\documentclass[prl,twocolumn,nofootinbib]{revtex4}
\usepackage{dcolumn}
\usepackage{amssymb}
\usepackage{bm}
\usepackage{enumerate}
\def\be{\begin{enumerate}}                     \def\ee{\end{enumerate}}
\def\beq{\begin{equation}}                     \def\eeq{\end{equation}}
\def\bea{\begin{eqnarray}}   \def\eea{\end{eqnarray}}

\def\3halfs{\textstyle{\frac{3}{2}}}                       

\def\ben{\begin{enumerate}}                   \def\een{\end{enumerate}}
\def\bitem{\begin{itemize}} \def\eitem{\end{itemize}}

\begin{document}
%\vspace{-1.0cm}
%--------------------------------------------------------------------
%\def\Universita{Universit\`a}
%--------------------------------------------------------------------
%\title{\bf \Large
\title{Ice Cube Observed PeV Neutrinos from AGN Cores}
%--------------------------------------------------------------------
\author{Floyd W. Stecker} \affiliation{NASA Goddard Space Flight Center \\
%\email[]{Floyd.W.Stecker@nasa.gov}
%\affiliation
Greenbelt, MD 20771, USA}

%-------------------------------------------------------------------
%\date{{\small Version 1, 20 October 2005; {\LaTeX-ed \today}}}
%--------------------------------------------------------------------
\begin{abstract}

I show that the high energy neutrino flux predicted to arise from AGN cores 
can explain the PeV neutrinos detected by Ice Cube without conflicting
with the constraints from the observed extragalactic cosmic ray and $\gamma$-ray 
backgrounds.
\end{abstract}
\maketitle
%--------------------------------------------------------------------
%\vfill
%%%%%%%%%%%%%%%%%%%%%%%%%%%%%%%%%%%%%%%%%%%

Recently, the Ice Cube collaboration has reported the first observation
of cosmic 2 PeV energy neutrinos giving a signal $\sim 3\sigma$ above the
atmospheric background~\cite{aa13}. More recently 18 events $\sim$4$\sigma$ 
above the expected atmospheric background were reported with energies 
above 100 TeV. These neutrinos are likely to be of cosmic origin;
their angular distribution is consistent with isotropy.
The average spectral index for these neutrinos was approximately -2
over the energy range between $\sim$ 100 TeV and 2 PeV~\cite{ha13}.

In 1991 we proposed a model suggesting that very high energy neutrinos
could  be produced in  the cores  of active  galaxies (AGN) such as Seyfert
galaxies  \cite{sdss91}. Using that  model, we  gave estimates  of the
flux  and  spectrum  of  high  energy neutrinos  to  be  expected. In
light of subsequent AGN observations and the discovery of neutrino 
oscillations the flux estimates for this model were revised downward
\cite{s05}, although the shape of the predicted neutrino spectrum
remained unchanged.

The new estimate was obtained by lowering the flux shown in
the Figure in Ref.~\cite{sdss91E} by a factor of 20.
This rescaling gives a value for the $\nu_{\mu}$ flux at 100 TeV
of $E_{\nu}^2\Phi (E_{\nu}) \sim 10^{-8}$ GeV cm$^{-2}$s$^{-1}$sr$^{-1}$
and a flux of $\sim 5.6 \times 10^{-8}$ GeV cm$^{-2}$s$^{-1}$sr$^{-1}$ at
$\sim 1$ PeV. The peak flux in these units occurs at an energy $\sim$1 PeV

In our model protons are accelerated by shocks in the cores of AGN 
in the vicinity of the black hole accretion disk (see, e.g., Ref. \cite{ke86}). 
Being trapped by the magnetic field, they lose energy dramatically 
by interactions with the dense photon field of the "big blue bump" of
thermal emission from the accretion disk (see, e.g., Ref. \cite{laor90})
which is optically thick to protons~\cite{sdss91}.
The primary interactions are those from photomeson production. The
primary neutrino producing channel, which occurs near threshold, is
 \begin{equation}
\label{photopion}
\gamma + p \rightarrow  \Delta^+ \rightarrow n + \pi^+.
\end{equation}
giving the pion an energy roughly $\approx 0.2 $ of that
of the emitting protons~\cite{st68}. 
Since the secondary $\gamma$-rays from $\Delta^+ \rightarrow p + \pi^0$
followed by $\pi^0 \rightarrow 2\gamma$ lose energy in the source from
electron-positron pair production and the protons in our model do not reach
ultrahigh energies, losing energy in the source,
there is no conflict with ultrahigh energy cosmic ray (UHECR) or extragalactic
$\gamma$-ray background constraints~\cite{ro13}. Full details of the model may be
found in Ref. \cite{sdss91}, Ref. \cite{s05} and references therein.

The important decays leading to $\nu$ production are $\pi^+ \rightarrow \mu^+ + \nu_{\mu}$
followed by $\mu^+ \rightarrow \bar{\nu_{\mu}} + \nu_e + e^+$, with all leptons carrying
off about 1/4 of the pion energy.  Thus in our model
we expect that with an assumed power-law proton spectrum from shock acceleration,
followed by photomeson production, there will be both more and higher energy
$\nu_e$'s produced than $\bar{\nu_e}$'s, reducing the effect of Glashow
resonance production at 6.3 PeV \cite{sg60} in the detector (see also Ref.~\cite{ro13}).

Given the effective area of Ice Cube for an isotropic flux of $\nu_e$'s at $\sim$1 PeV
of 5 m$^2$ and an exposure time of 615.9 days~\cite{aa13}, and giving a total predicted $\nu$
flux at $\sim$1 PeV of $\sim 6 \times 10^{-14}$ (cm$^2$ s sr)$^{-1}$,
we would predict that Ice Cube should see a total of $\sim$6 neutrino induced events. 

The main uncertainty in the magnitude of the modeled flux is from the 
uncertainty in the number of AGN per Mpc$^3$. The neutrino spectrum
was predicted to be proportional to $E^{-2.1}$ between 1 and 10 PeV under
the assumption that the proton spectrum has an $E^{-2}$ dependence from
shock acceleration. The average slope of the neutrino spectrum was 
predicted to be $\sim 3.1$ between 1 and 100 PeV. Protons were assumed to
be accelerated up to a maximum energy of $2.5 \times 10^4$ PeV. It is, of course,
possible that the average proton spectrum can be steeper than $E^{-2}$ and that
the maximum energy to which protons are accelerated is less than $2.5 \times 10^4$ PeV. 
Since the sources are extragalactic, we expect that the observed neutrinos will have
a roughly isotropic angular distribution on the sky.

Given the uncertainties in the model parameters, the general agreement with the AGN
core model is significant, particularly the prediction of a peak neutrino energy flux
at $\sim$ 1 PeV. I conclude that, given the present Ice Cube results, AGN cores may naturally account for the implied $\nu$ flux and angular distribution without violating constraints from $\gamma$-ray background and UHECR fluxes. More Ice Cube data should soon be forthcoming. 

Acknowledgment: I thank Francis Halzen for helpful comments.

%----------------------------------------------------------------

%---------------------------------------------------------------

\begin{thebibliography}{99}


\bibitem{aa13} M. G. Aartsen et al., arXiv:1304.5356.

\bibitem{ha13} F. Halzen, {\it Proc. 33rd Intl. Cosmic Ray Conf.,
Rio de Janiero}, 2013.

\bibitem{sdss91} F. W. Stecker, C. Done, M.H. Salamon and P. Sommers,
Phys.\ Rev.\ Lett.\  {\bf 66}, 2697 (1991), 

\bibitem{sdss91E} F. W. Stecker, Phys.\ Rev.\ Lett.\  
{\bf 69}, 2738 (1992).

\bibitem{s05} F. W. Stecker, Phys.\ Rev.\ D \  {\bf 72}, 107301 (2005).

\bibitem{ke86} D. Kazanas and D. C. Ellison, Astrophys. J. {\bf 304}, 178 (1986).

\bibitem{laor90} A. Laor, Mon. Not. Royal Astr. Soc. {\bf 246}, 369 (1990).

\bibitem{st68} F. W. Stecker, Phys.\ Rev.\ Lett.\ {\bf 21}, 1016 (1968).

\bibitem{ro13} E. Roulet et al., J. Cosmol. Astropart. Phys. 2013/01/028 (2013).

\bibitem{sg60} S. L. Glashow, Phys. Rev. {\bf 118}, 316 (1960).

%---------------------------------------------------------------
\end{thebibliography}
\end{document}